\documentclass[12pt]{article}
\usepackage{amsmath}
\usepackage{amsfonts}
\usepackage{amssymb}
\usepackage{amsthm}
\usepackage[dvips]{color}
\usepackage{graphicx}
\usepackage{calc}
\usepackage{psfrag}
\def\Bmp#1{ \begin{minipage}{#1} }
\def\Bmpc#1{ \begin{minipage}[c]{#1} }
\def\Bmpt#1{ \begin{minipage}[t]{#1} }
\def\Bmpb#1{ \begin{minipage}[b]{#1} }
\def\Emp{ \end{minipage} }

\def\L{{\mathcal{L}}}

\def\O{{\mathcal{O}}}

\def\RR{{\mathbb{R}}}
\def\CC{{\mathbb{C}}}
\def\ZZ{{\mathbb{Z}}}

\def\a{{\bf a}}
\def\b{{\bf b}}

\def\n{{\bf n}}

\def\x{{\bf x}}

\def\y{{\bf y}}
\def\0{{\bf 0}}
\def\Z{{\bf Z}}
\def\t{{\bf t}}
\def\V{{\bf V}}
\def\A{{\bf A}}
\def\B{{\bf B}}
\def\C{{\bf C}}
\def\D{{\bf D}}

\def\bnabla{\boldsymbol{\nabla}}

\def\Dpartial#1#2{ {\partial #1 \over \partial #2} }

\newtheorem{lemma}{Lemma}

\begin{document}

%\vspace*{-1.0cm}
\title{A Framework for Linear Stability Analysis of Finite-Area Vortices}
\author{Alan Elcrat$^1$ and Bartosz Protas$^2$
\\ \\
$^1$Department of Mathematics, \\
Wichita State University, Wichita, KS, USA 
\\ 
$^2$Department of Mathematics \& Statistics, \\
McMaster University, Hamilton, ON, Canada }

\date{\today}
\maketitle

\begin{abstract}
In this investigation we revisit the question of the linear
  stability analysis of 2D steady Euler flows characterized by the
  presence of compact regions with constant vorticity embedded in a
  potential flow. We give a {complete} derivation of the
  linearized perturbation equation which, recognizing that the
  underlying equilibrium problem is of the free-boundary type, is done
  systematically using methods of the shape-differential calculus.
  Particular attention is given to the proper linearization of the
  contour integrals describing vortex induction.  The thus obtained
  perturbation equation is validated by analytically deducing from it
  the stability analyses of the circular vortex, originally due to
  Kelvin (1880), and of the elliptic vortex, originally due to Love
  (1893), as special cases. We also propose and validate a
  spectrally-accurate numerical approach to the solution of the
  stability problem for vortices of general shape in which all
  singular integrals are evaluated analytically.

\end{abstract}

% \keywords{
%Vortex Dynamics; Linear Stability; 
%Free-Boundary Problems; Shape Calculus
%}

%\classification{...}

\maketitle

%\clearpage

\section{Introduction}
\label{sec:intro}

In this work we are interested in the linear stability of solutions of
the two-dimensional (2D) Euler equations which are steady in the
appropriate frame of reference and feature compact regions with
constant vorticity embedded in an otherwise potential flow. Such
solutions arise as inviscid limits of actual viscous flows, and
therefore find numerous applications in different areas of fluid
mechanics. The first linear stability analysis of such a flow is
attributed to Kelvin \cite{k80}, see also \cite{l32,b67,s92}, who
identified the dispersion relation of neutrally-stable travelling
waves with $m$-fold symmetry moving along the perimeter of a circular
vortex. This problem was revisited by Baker \cite{b90} who used a
contour dynamics approach involving integrals with singular kernels
defined on the vortex boundary. The stability of a rotating ellipse,
the so-called Kirchhoff vortex \cite{k76}, was analyzed in the seminal
study by Love \cite{l93}. The effect of an ambient strain field was
investigated by Moore \& Saffman \cite{ms71}, whereas Mitchell \&
Rossi \cite{mr08} explored the relation between the linear
instabilities and the long-term nonlinear evolution of elliptic
vortices. In this context we also mention the analytical study of Guo
et al.~\cite{ghs04} where an integro-differential perturbation
equation similar to ours is proposed for the case of the elliptic
vortex.  The linear stability of more complex vortex configurations,
such as polygonal arrays of corotating vortices and translating vortex
pairs, was investigated by Dritschel \cite{d85,d90,d95}, see also
Dritschel \& Legras \cite{dl91}. A noteworthy feature of their
approach is that they also used a {\em continuous} perturbation
equation independent of a particular discretization employed to
obtain the equilibrium solution. Problems related to the stability of
polygonal vortex arrays were also studied by Dhanak \cite{d92}.  The
linear stability of the so-called V-states \cite{woz84}, corotating
vortex patches and infinite periodic arrays of vortices was
investigated in detail by Kamm \cite{k87}, see as well \cite{s92}.
This approach was based on the representation of the solution in terms
of the Schwarz function, and required discretization and numerical
differentiation in order to obtain the perturbation equation. A
discrete form of the perturbation equation was also used by Elcrat et
al.~\cite{efm05} in their investigation of the linear stability of
vortices in a symmetric equilibrium with a circular cylinder and a
free stream at infinity.  There have also been a number of analytical
investigations concerning the stability, both in the linear and
nonlinear setting, of flows obtained as perturbations of the Rankine
vortex \cite{b82,bl82,t85,t92,w86,w98}. Another family of approaches
is based on variational energy arguments going back to Kelvin. They
were initially investigated in \cite{d85,ss80,d88,cw03,fm08}, and were
more recently pursued by Luzzatto-Fegiz \& Williamson
\cite{fw10a,fw10b,fw11a,fw11b,fw12a,fw12b}. They rely on global
properties of the excess energy vs.~velocity impulse diagrams and
provide partial information about the linear stability properties
without the need to actually perform a full linear stability analysis.

To the best of our knowledge, there is no {complete} derivation
and validation of the stability equations for two-dimensional (2D)
vortex patches available in the literature. The main goal of this
paper is thus to fill this gap by proposing a systematic approach to
the study of the linear stability of vortices which is quite general
in the sense that it is based on the continuous, rather than discrete,
formulation of the governing equations and a rigorous treatment of
boundary deformations. {This aspect, namely, that linearization
  of differential and integral expressions with respect to
  perturbations of the shape of the domain requires special treatment
  with methods other than the traditional calculus of variations does
  not seem to have received much attention in the vortex dynamics
  literature.}  Therefore, the proposed approach may serve as a
template for studying more complicated problems such as the stability
of vortex equilibria in three dimensions (3D). Recognizing the Euler
equation describing finite-area vortices as a {\em free-boundary}
problem, we apply methods of the shape-differential calculus to derive
a perturbation equation in the form of an integro-differential
equation defined on the vortex boundary. We emphasize that, in
contrast to \cite{k87,efm05}, numerical differentiation is not
required at any stage of the derivation, but only in the evaluation of
the coefficients of the perturbation equation. In this sense, the
formulation we propose is a contour-dynamics analogue of the
Orr-Sommerfeld equation used to study the stability of viscous
parallel flows \cite{dr04}. In this work we also demonstrate using
analytical calculations how one can reproduce the dispersion relations
obtained by Kelvin in the stability analysis of the Rankine vortex
\cite{k80} and by Love in the stability analysis of the elliptic
vortex \cite{l93} from our general perturbation equation.
Generalization of our approach to the stability analysis of more
complicated vortex flows, including 3D axisymmetric configurations, is
deferred to future research.

We are interested in developing a general approach to analyzing the
linear stability of 2D steady-state solutions of the Euler equation
which is usually written in the following form \cite{mb02}
\begin{subequations}
\label{eq:euler}
\begin{alignat}{2}
\Delta \psi & = F( \psi) \qquad && \textrm{in} \ \Omega, \label{eq:euler_a} \\
\psi & = \psi_b && \textrm{on} \ \partial\Omega, \label{eq:euler_b}
\end{alignat}
\end{subequations}
where $\psi$ is the streamfunction, $\Omega \subseteq \RR^2$ is the
flow domain, whereas $\psi_b$ represents the boundary condition for
the streamfunction. The function $F \: : \: \RR \rightarrow \RR$ is a
priori undetermined and any choice of $F$ will yield a steady solution
of the 2D Euler equation. One choice of the function $F$, motivated by
the flow models arising from the Prandtl-Batchelor theory \cite{ch09},
is
\begin{equation}
F(\psi) = -\omega\, H(\psi_0 - \psi),
\label{eq:F}
\end{equation}
where $H(\cdot)$ is the Heaviside function, whereas $\omega,\psi_0 \in
\RR$ are, respectively, the vorticity inside the vortex and the value
of the streamfunction at its boundary. The profile given in
\eqref{eq:F} corresponds to finite-area vortices with constant
vorticity $\omega$ embedded in an otherwise potential flow. This
formulation can be made more general by including in \eqref{eq:F} a
term proportional to Dirac delta function representing a vortex sheet
on the vortex boundary \cite{wmz06}, but we do not consider this in
the present study. A salient feature of problem
\eqref{eq:euler}--\eqref{eq:F} is that the shape of the vortex region,
which we will denote $A$, is a priori unknown and therefore must be
determined as a part of the solution to the problem. System
\eqref{eq:euler}--\eqref{eq:F} is thus a {\em free-boundary} problem,
and in order to emphasize this fact which will play a central role in
our analysis, we rewrite it in the following form which makes the
free-boundary property more evident
\begin{subequations}
\label{eq:eulerFB}
\begin{alignat}{2}
& \Delta \psi_1 = - \omega & & \textrm{in} \ A, \label{eq:eulerFBa} \\
& \Delta \psi_2 = 0 & & \textrm{in} \ \Omega \backslash \overline{A}, \label{eq:eulerFBb} \\
& \psi_1 = \psi_2 = \psi_0 \hspace*{1.0cm} & & \textrm{on} \ \partial A, \label{eq:eulerFBc} \\
& \Dpartial{\psi_1}{n} = \Dpartial{\psi_2}{n} & & \textrm{on} \ \partial A, \label{eq:eulerFBd} \\
& \psi_2 = \psi_b & & \textrm{on} \ \partial\Omega, \label{eq:eulerFBe}
\end{alignat}
\end{subequations}
where $\psi_1 \triangleq \psi|_{A}$ and $\psi_2 \triangleq
\psi|_{\Omega \backslash \overline{A}}$ are the restrictions of the
streamfunction $\psi$ to, respectively, the rotational and
irrotational part of the flow domain and $\n$ is the unit vector
normal to the vortex boundary (``$\triangleq$'' means ``equal to by
definition''). Solutions of system \eqref{eq:eulerFB} depend on two
parameters $\omega$ and $\psi_0$ or, equivalently, the vortex area
$|A| \triangleq \int_{\Omega} H(\psi_0 - \psi)\, d\Omega$ and its
circulation $\gamma \triangleq \omega |A|$. There is evidence that by
continuing the solutions of \eqref{eq:eulerFB} corresponding to
different flow configurations such that $\gamma = Const$ and $|A|
\rightarrow 0$ one can obtain the corresponding point-vortex
equilibria \cite{t83a,t83b,gipz09a}. In agreement with earlier studies
\cite{k87,d95}, we will assume in this investigation that both
parameters $|A|$ and $\gamma$ are fixed, so that the perturbations
will be allowed to modify the shape of the vortex region $A$ leaving
its area and circulation unchanged. In other words, no vorticity is
created by the perturbations.  Derivation of the Jacobian, needed for
the linear stability analysis of a given problem, requires
differentiation of the governing equation with respect to the state
variables. In the present problem, since governing system
\eqref{eq:eulerFB} is of the free-boundary type, this is a rather
delicate issue, because the shape of the vortex $A$ can be regarded as
a ``state variable''. A proper way to deal with this problem is
therefore to use the {\em shape-differential calculus}
\cite{sz92,dz01a,hm03} which is a suite of mathematical techniques for
differentiation of various objects such as differential equations,
integrals, etc., defined on variable domains. Application of the
shape-differentiation methods makes it possible to derive a continuous
perturbation equation for vortices with very general shapes in a
rigorous and systematic manner which is the main contribution of this
work. A key element is a proper linearization of the contour
  dynamics equations with respect to arbitrary boundary
  perturbations. As regards derivations of the different
shape-differentiation identities we use in this study, the reader is
referred to monographs \cite{sz92,dz01a,hm03}, and to the review paper
\cite{ss10} for a survey of applications of the shape calculus to
different problems in computational fluid dynamics.  The structure of
the paper is as follows: in the next Section we derive a general form
of the perturbation equation, in Sections \ref{sec:circular} and
\ref{sec:elliptic} we use this perturbation equation to obtain the
dispersion relations for the circular and elliptic vortices, comments
about an accurate numerical technique for the solution of the
perturbation equation are presented in Section \ref{sec:numer}
together with some computational results, whereas conclusions
are deferred to Section \ref{sec:final}. Some technical results
  are collected in Appendix.

\section{Derivation of Perturbation Equation}
\label{sec:pert}

\begin{figure}
\begin{center}
%\psfrag{dA}{$\partial A$}
%\psfrag{w}{$\tilde{z}^{\epsilon}$}
\includegraphics[width=0.6\textwidth]{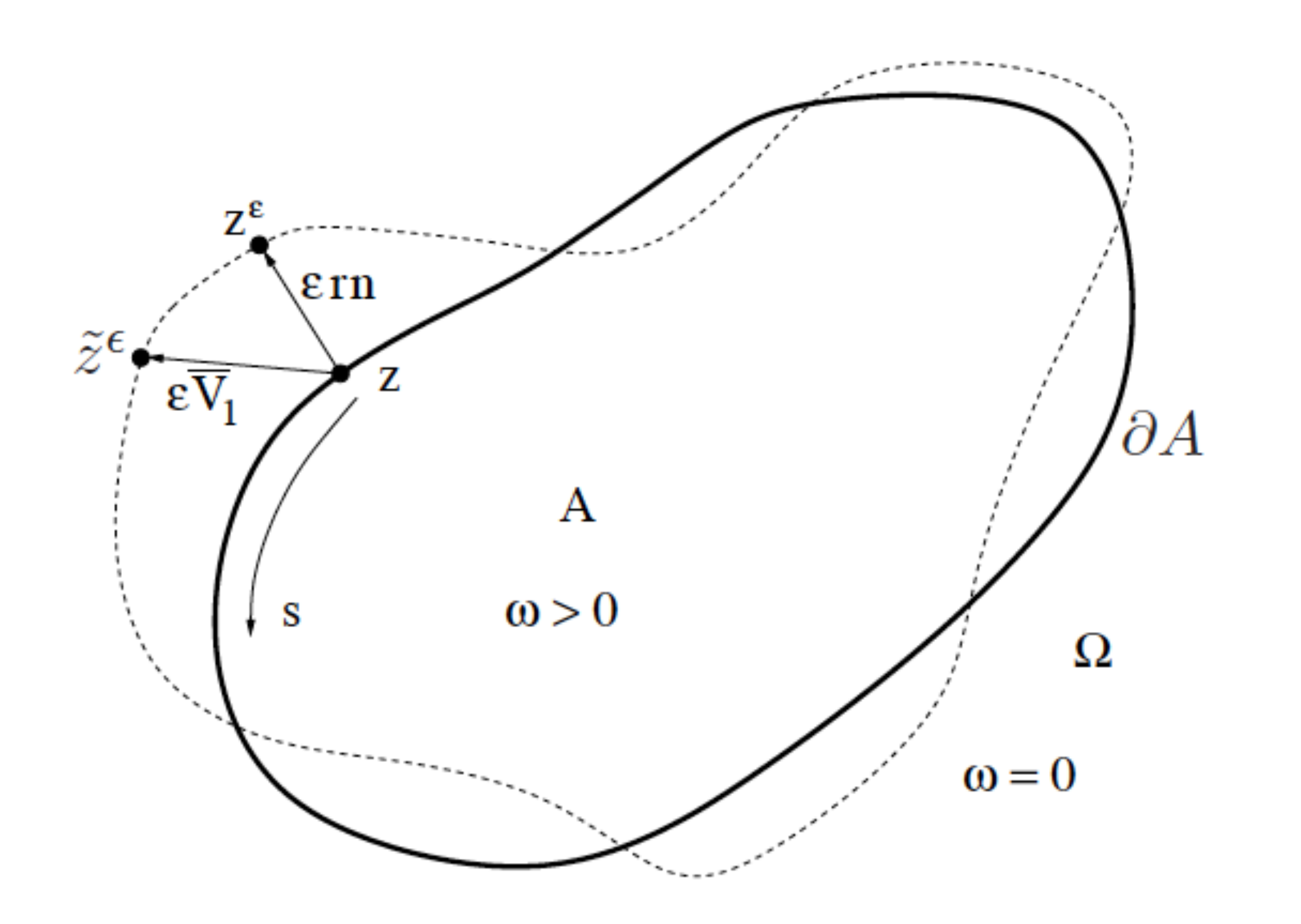}
\caption{Schematic illustrating (solid line) the equilibrium vortex
  boundary $\partial A$ and (dashed line) its perturbation. The points
  $\tilde{z}^{\epsilon}$ and $z^{\epsilon}$ represent the images of
  the point $z$ under, respectively, the Lagrangian and ``geometric''
  displacement.}
\label{fig:patch}
\end{center}
\end{figure}

In this Section we develop a systematic approach to derivation of the
perturbation equation characterizing the stability of steady vortex
configurations. For brevity of notation, in what follows we will use
the complex number representation of vector quantities. This is
indicated by using ordinary letters for vector quantities (e.g.,
$n \in \CC$ for $\n\in\RR^2$ and similarly for other
vector quantities). Juxtaposition (i.e., $z_1z_2$ for some $z_1$, $z_2
\in \CC$) indicates complex multiplication of $z_1$ and $z_2$.
However, when we need equations written in the vector form, as we will
in referencing results from shape calculus, we will denote vector
quantities in bold face and inner products of vectors using the usual
dot notation. Thus, for $\a, \b\in \RR^2$ vectors and $a, b \in \CC$
the corresponding complex numbers, we have the following identity $\a
\cdot \b = \Re[a \,\overline{b}]$, where the overbar denotes
  complex conjugation. To fix attention, we will focus on the
simplest, yet generic, case of a single vortex region $A$ in
equilibrium with an external velocity field $U = u - iv$ which may
represent a free stream, the influence of other vortices, or may
result from transformation of the problem to a moving (i.e.,
translating or rotating) coordinate system in which a relative
equilibrium exists. We will consider a point $z=x+iy$ on the vortex
boundary $\partial A$, see Figure \ref{fig:patch}, which is assumed to
have counterclockwise orientation. Using techniques of vortex dynamics
\cite{s92}, the velocity of a particle located at the point $z$ can be
expressed in the fixed frame of reference as follows \footnote{We
  remark that the same integral appears in \cite{s92,Pullin81} with
  the opposite sign.  This is due to a different orientation of the
  contour in these studies.}
\begin{equation}
\frac{d\overline{z}}{d\tau} = V(\tau,z) =  \frac{\omega}{4\pi} \oint_{\partial A} 
\frac{\overline{z} - \overline{w}}{z - w}\, dw + U(\tau,z),
\label{eq:V}
\end{equation}
where the integral on the right-hand side (RHS) has a removable
singularity ($w$ is a complex integration variable). Starting from
relation \eqref{eq:V}, one can pursue our objective with one of the
two alternative approaches, namely (Figure \ref{fig:patch}):
\begin{itemize}
\item
{\it Lagrangian} approach in which one follows the trajectory of a
fluid particle on the vortex boundary; we note that displacement of a
material particle in general also involves a component tangential to the
vortex boundary which does not lead to deformations of the contour
\cite{b90},

\item
{\it geometric} approach in which one considers displacements of the
points on the boundary in the direction normal to the boundary only
\cite{d95}.
\end{itemize}
Since the Lagrangian approach involves additional (tangential) degrees
of freedom, the stability results obtained with the geometric approach
represent a subset of the results obtained with the Lagrangian
approach. While both approaches can be pursued using methods of
shape-differentiation, in this work we will focus on the geometric
approach as it leads to somewhat simpler calculations.

As a first step, we transform relation \eqref{eq:V} to a suitable
moving (i.e., rotating or translating) coordinate system,
characterized by the rigid-body velocity $V_0(z)$, in which a relative
equilibrium {is attained}. Denoting $\zeta(\tau)$ the position of
the point $z(\tau)$ in this moving coordinate system, the left-hand
side (LHS) of \eqref{eq:V} becomes $\frac{d\overline{z}}{d\tau} =
V_0(z) + \frac{d\overline{\zeta}}{d\tau}$, so that relation
\eqref{eq:V} can rewritten as
\begin{equation}
\frac{d\overline{\zeta}}{d\tau} =  \frac{\omega}{4\pi} \oint_{\partial A} 
\frac{\overline{z} - \overline{w}}{z - w}\, dw + U(\tau,z) - V_0(z) \triangleq V_1(\tau,z).
\label{eq:V1}
\end{equation}
In equation \eqref{eq:V1} and hereafter we use both $z(\tau)$ and
$\zeta(\tau)$, and note that, since $\zeta(0) = z(0)$, either
coordinate can be used on the RHS of \eqref{eq:V1} to derive the
perturbation equation. Let $n = n_x + in_y$ be the unit vector normal
to the boundary $\partial A$ and pointing outside of $A$.  The
relative equilibrium is then characterized by the condition
\begin{equation}
\Re\left[ n \, \frac{d\overline{\zeta}}{d\tau} \right] = 
\Re\left[ n \, \left( \frac{\omega}{4\pi} \oint_{\partial A} 
\frac{\overline{z} - \overline{w}}{z - w}\, dw + U(\tau,z) - V_0(z) \right) \right] = 0
\label{eq:rel_eq}
\end{equation}
which, for given vorticity $\omega$ and vortex area $|A|$, can
interpreted as a condition on the shape of the vortex boundary
$\partial A$. Next, we introduce a parametrization of the contour $z =
z(\tau,s) \in \partial A$, where $\tau \in \RR^+$ is the time, such
that $s \in [0,L]$ in which $L$ denotes the length of the boundary
$\partial A$. We will also adopt the convention that the superscript
$\epsilon$, where $0 < \epsilon \ll 1$, will denote quantities
corresponding to the perturbed boundary, so that $\zeta =
\zeta^{\epsilon}|_{\epsilon = 0}$ and $n = n^{\epsilon}|_{\epsilon =
  0}$ are the quantities corresponding to the (relative) equilibrium.
Then, points on the perturbed vortex boundary can be represented as
follows
\begin{equation}
\zeta^{\epsilon}(\tau,s) = \zeta(s) + \epsilon\, r(\tau,s) \, n(s),
\label{eq:z}
\end{equation}
where $r(t,s)$ represents the ``shape'' of the perturbation. We note
that, without affecting the final result, $n(s)$ in the last term in
\eqref{eq:z} could be replaced with its perturbed counterpart
$n^{\epsilon}(\tau,s)$. Using equation \eqref{eq:V} we thus deduce
\begin{equation}
\Re\left[n^\epsilon \frac{d\overline{\zeta^{\epsilon}}}{d\tau}\right] 
=  \Re\left[n^\epsilon \frac{\omega}{4\pi} \oint_{\partial A^{\epsilon}} 
\frac{\overline{z^{\epsilon}} - \overline{w}}{z^{\epsilon} - w}\, dw\right] 
+ \Re\left[n^\epsilon \left( U(z^{\epsilon}) - V_0(z^{\epsilon})\right)\right],
\label{eq:Vn}
\end{equation}
from which equilibrium condition \eqref{eq:rel_eq} is obtained by
setting $\epsilon = 0$.  In \eqref{eq:Vn} the integral operator is
defined in terms of a contour integral. In order to be able to use
shape calculus formulas, we need to write such integrals as integrals
with respect to the arc length.  If $F$ is a complex-valued function
and $\Gamma$ a closed $C^1$ curve, we have
\begin{equation}
  \oint_{\Gamma} F(w)\,dw = \oint_{\Gamma} F(w) t \, |dw|,
\label{eq:dw}
\end{equation}
where $t$ is the unit tangent vector in the counterclockwise direction.

The perturbation equation is obtained by linearizing equation
\eqref{eq:Vn} around the equilibrium configuration characterized by
\eqref{eq:rel_eq} which is equivalent to expanding \eqref{eq:Vn} in
powers of $\epsilon$ and retaining the first-order terms. Since
equation \eqref{eq:Vn} involves perturbed quantities defined on the
perturbed vortex boundary $\partial A^{\epsilon}$, the proper way to
obtain this linearization is using methods of the shape-differential
calculus \cite{dz01a}.  Differentiating the LHS in \eqref{eq:Vn} and
setting $\epsilon = 0 $ we thus obtain
\begin{equation}
\frac{d}{d\epsilon}\left\{ \Re\left[ n^\epsilon \frac{d(\overline{\zeta} + 
\epsilon \,r\, \overline{n})}{dt} \right]\right\}\bigg|_{\epsilon = 0} = 
\Re\left[\frac{d n^\epsilon}{d \epsilon}\bigg|_{\epsilon = 0} 
\frac{d\overline{\zeta}}{dt} + n \frac{ d (r\, \overline{n}) }{dt}\right].
\label{eq:dlhs}
\end{equation}
It can be shown using shape calculus \cite{dz01a} that
\begin{equation}
\frac{d \n^\epsilon}{d\epsilon}\bigg|_{\epsilon = 0} =  - \Dpartial{r}{s} \t.
\label{eq:dn}
\end{equation}
Noting also that $\frac{d\n}{d\tau}\big|_{\tau = 0} = -
\left(\bnabla_{\Gamma} \V \right)^T \n$, where $\bnabla_{\Gamma}$
denotes the tangential gradient \cite{dz01a}, has only a component
tangential to the boundary $\partial A$ and expanding
$\frac{dr}{d\tau} = \Dpartial{r}{\tau} + V_{1t} \Dpartial{r}{s}$,
where $V_{1t} \triangleq \Re(V_1 \, t )$ and $\Dpartial{}{s}$ is the
derivative with respect to the arc length coordinate, we can rewrite
\eqref{eq:dlhs} as follows
\begin{equation}
  \frac{d}{d\epsilon}\left\{ \Re\left[ n^\epsilon \frac{d(\overline{\zeta} + 
        \epsilon \,r\, \overline{n})}{dt} \right]\right\}\bigg|_{\epsilon = 0} = 
  \Re\left[ - \Dpartial{r}{s} \, t \, \frac{d\overline{\zeta}}{d\tau} + 
\frac{dr}{d\tau} \, n \, \overline{n} + r \, n \, \frac{d\overline{n}}{d\tau} \right] 
= \Dpartial{r}{\tau}.
\label{eq:dlhs2}
\end{equation}

As regards the RHS in \eqref{eq:Vn}, we note that the dependence on
the perturbation $\epsilon r n$ is more complicated here, as the
perturbation also affects the {\em shape} of the contour $\partial
A^{\epsilon}$ on which the integral is defined. {To evaluate
  these expressions, we need the concept of the ``shape derivative''.
  Given a function $f \; : \; \Omega \rightarrow \RR$ and a
  perturbation vector field $\Z \; : \; \Omega \rightarrow \RR^2$ such
  that $\Z|_{\partial\Omega} = \0$, the shape derivative $f'$ of $f$
  with respect to perturbation $\Z$ is defined through the identity
  \cite{hm03}
\begin{equation}
\frac{d}{d\epsilon} f(\x+\epsilon\,\Z(\x))\big|_{\epsilon = 0} = f'(\x) + \left(\bnabla f(\x)\right)\cdot\Z(\x),
\qquad \forall_{\x \in \Omega}.
\label{eq:df}
\end{equation}
In our problem
\begin{equation}
f(z) =  \frac{\omega}{4\pi} \oint_{\partial A} 
\frac{\overline{z} - \overline{w}}{z - w}\, dw + U(z) - V_0(z)
\label{eq:f} 
\end{equation}
and in order to apply relation \eqref{eq:df} we need to translate the
vector operations into complex notation}. This can be done with the
help of the following lemma.
\begin{lemma}
  For $f(w,\overline{w})$ a complex-valued function, $\Z = [Z_1,Z_2]$
  and $Z=Z_1+iZ_2$, the directional derivative of $f$ in the direction
  $\Z$ can be expressed as
\begin{equation}
 \bnabla f \cdot \Z = \frac{\partial f}{\partial w}Z + 
\frac{\partial f}{\partial \overline{w}}\overline{Z},
\end{equation}
where $\frac{\partial}{\partial w} \triangleq \frac{1}{2}(\frac{\partial}{\partial x} - 
i\frac{\partial}{\partial y})$ and $\frac{\partial}{\partial \overline{w}} \triangleq \frac{1}{2}(
\frac{\partial}{\partial x} + i\frac{\partial}{\partial y})$ when $w=x+iy$.
\end{lemma}
We can then write for the right hand side of \eqref{eq:Vn}
\begin{equation}
\begin{aligned}
\frac{d}{d\epsilon} \Big\{ \Re\left[ n^\epsilon \, f^\epsilon \right] \Big\}\Big|_{\epsilon = 0} 
& = \frac{d}{d\epsilon} \left\{ \Re\left[ n^\epsilon  \frac{\omega}{4\pi}
\oint_{\partial A^\epsilon}\frac{\overline{z^\epsilon} - \overline{w}}{z^\epsilon - w}\, dw + 
n^\epsilon ( U(z^\epsilon) - V_0(z^\epsilon))\right]\right\}\bigg|_{\epsilon = 0} \\
& = \Re\left[\frac{d n^\epsilon}{d \epsilon}\bigg|_{\epsilon = 0} f +
n \left(f' + r\left(n\,\frac{\partial}{\partial z}
+ \overline{n}\,\frac{\partial}{\partial \overline{z}}\right)f\right)\right],
\end{aligned}
\label{eq:drhs}
\end{equation}
where we assume $\Z = r\,\n$ in \eqref{eq:df}.

As concerns the shape derivative $f'$ in \eqref{eq:drhs}, since the
reference velocity of the moving coordinate system $V_0(z)$ does not
depend on the perturbation, we have $[V_0(z)]' \equiv 0$. For the
integral term we need to use the following shape-calculus formula
\cite[page 354, equation (4.17)]{dz01a}
\begin{equation}
\left[ \oint_ {\partial A^\epsilon} g\, |dw|\right]' 
= \oint_{\partial A} \left[ g' + \left(\Dpartial{g}{n} + \kappa \, g\right)(\Z\cdot\n)\right]\,|dw|,
\label{eq:dJ}
\end{equation}
where $g \: : \: \Omega \rightarrow \RR$ is a differentiable function
($\partial A^\epsilon \subset \Omega$), $g'$ its shape derivative and
$\kappa$ is the curvature of the contour $\partial A = \Gamma$. While
in the standard texts \cite{dz01a,hm03} formula \eqref{eq:dJ} is
derived for real-valued functions $g$, an extension to complex-valued
integrals is straightforward. For us, then, cf.~\eqref{eq:V} and
\eqref{eq:dw}
\begin{equation}
g(s,q) = \frac{\omega}{4\pi} \frac{\overline{z^\epsilon}(s) - \overline{w}(q)}{z^\epsilon(s) - w(q)} t,
\qquad s,q \in [0,L].
\label{eq:g}
\end{equation}
Hereafter we will adopt the convention that the subscripts $z$ and $w$
on different symbols, e.g.,~$n$, $t$, or $r$, will indicate where the
corresponding quantity is evaluated. As regards the shape
derivative $g'$ in \eqref{eq:dJ}, we note that $g$ depends on the
perturbation $\Z$ through the independent variables only, so that by
\eqref{eq:df} we have
$\frac{dg(\epsilon,\Z)}{d\epsilon}\big|_{\epsilon = 0} = \bnabla g
\cdot\Z$ and therefore $g' \equiv 0$. Finally, assembling
\eqref{eq:dlhs2} and \eqref{eq:drhs}, using \eqref{eq:dJ}, and
performing the differentiations required in \eqref{eq:drhs} and
\eqref{eq:dJ}, we obtain after rearranging terms the following
perturbation equation
\begin{equation}
\begin{aligned}
\Dpartial{r}{\tau} =  - V_{1t} \, \Dpartial{r}{s} +   
& \Re\left[ n_z \, U'(z) + r_z \, n_z \left( \frac{dU}{dz} - \frac{dV_0}{dz}\right)\right] \\
- \frac{\omega}{4\pi} \Re\Bigg[ 
    n_z  & \oint_{\partial A} \frac{\overline{z} - \overline{w}}{z - w} \left(i\,\frac{dr}{ds} - \kappa\, r_w\right)\, t_w\,|dw| \\
 - n_z & \oint_{\partial A}\frac{(\overline{z}-\overline{w})n_w-(z-w)\overline{n}_w}{(z-w)^2}\,r_w \, t_w \,|dw| \\
 + n_z \, r_z & \oint_{\partial A} \frac{n_z(\overline{z}-\overline{w}) - \overline{n}_z (z-w)}
{(z-w)^2}\, t_w \,|dw|\Bigg]\\
\end{aligned}
\label{eq:dr}
\end{equation}
describing the linear evolution of the local amplitude $r(\tau,s)$ of
the vortex boundary perturbation in the normal direction
[cf.~\eqref{eq:z}]. In the above we have also used the identity $t=i
\, n$ which follows from the counter-clockwise orientation of
$\partial A$.  In combination with an initial condition
\begin{equation}
r(0,s) = r_0(s), \qquad s \in [0,L],
\label{eq:r0}
\end{equation}
system \eqref{eq:dr}--\eqref{eq:r0} should be interpreted as an
initial-value problem in which the solution $r(t,s)$ is subject to the
periodic boundary condition $r(\tau,0) = r(\tau,L)$, $\forall \tau \ge
0$. We also add that some of the integrals in \eqref{eq:dr} are to be
understood in the principal value sense. We remark that, although some
terms are present in both equations, equation \eqref{eq:dr} appears
different from the perturbation equation obtained by Dritschel
\cite{d95}. The assumption that, to the leading order in $\epsilon$,
the perturbations do not change the vortex area, i.e., $|A^\epsilon| =
Const + \O(\epsilon^2)$, leads to a restriction on the
admissible initial perturbations $r_0$. Shape-differentiating the
expression for the vortex area we obtain
\begin{equation}
\left[ \int_{\Omega} H(\psi_0 - \psi) \,d\Omega\right]' = \left[ \int_{A^\epsilon}\, d\Omega\right]' = 
\oint_{\partial A} r_0 \overline{n} n \, ds = \oint_{\partial A} r_0 \, ds = 0
\label{eq:dA}
\end{equation}
which means that, in order to satisfy (at the initial instant of time)
the constant-area condition, we need to restrict attention to initial
perturbations $r_0$ with the vanishing mean with respect to the arc
length $s$. We add that, assuming constant vorticity $\omega = Const$,
condition \eqref{eq:dA} also ensures that the vortex circulation
$\gamma$ remains unchanged by perturbations. {Construction of
  perturbations satisfying condition \eqref{eq:dA} is discussed in
  Section \ref{sec:numer}, cf.~equation \eqref{eq:rn}.}  In the
absence of the constant-area condition \eqref{eq:dA}, our preceding
analysis would need to be slightly modified, as then the shape
derivative $g'$ would not vanish.

In the following two Sections we show how the
stability analyses of the circular and elliptic vortices can be
derived as special cases from general perturbation equation
\eqref{eq:dr}.

\section{Stability of Circular Vortex}
\label{sec:circular}

In this Section we demonstrate how the well-known results concerning
the stability of the Rankine vortex originally obtained by Kelvin
\cite{k80} and reviewed also in \cite{l32,b67,s92}, in particular, the
dispersion relation for the instability waves travelling along the
vortex boundary, can be analytically deduced directly from general
perturbation equation \eqref{eq:dr}. Even though these results are
classical, we go through these calculations in some detail, because
they will also play an important role in the stability calculations
for vortices with arbitrary shapes. More specifically, as will be
shown in Section \ref{sec:numer} of this work, numerical stability
computations for vortices with general shapes can be reduced to a
``perturbation'' of the corresponding analysis for the circular
(Rankine) vortex. The main advantage of this approach is that it
allows us to compute all singular (principal-value) integrals
analytically. For the Rankine vortex with radius $a$ we thus set $z(p)
= a e^{ip}$, $n(p) = e^{ip}$, and $\kappa = \frac{1}{a}$, where
$p=\frac{2 \pi}{L}s \in [0,2\pi]$.  As regards the moving coordinate
system, due to the rotational invariance of the problem, its velocity
$V_0(z)$ can be selected arbitrarily and without loss of generality we
choose $V_0(z) \equiv 0$ resulting in $V_{1t} = \frac{1}{2} \omega a$
in \eqref{eq:dr}. The perturbation equation then simplifies to
\begin{equation}
\Dpartial{r}{\tau} = -\frac{\omega }{2} \Dpartial{r}{p} + 
\frac{ \omega}{4\pi} \Re\left[i\int_0^{2\pi}\left(i\frac{\partial r}{\partial q}-r(q)\right)\,dq\right] + 
\frac{\omega}{4\pi} \Re\left[i\int_0^{2\pi}\frac{(e^{iq} + e^{ip}) r(q)}{e^{ip} - e^{iq}}\,dq\right],
\label{eq:dr0}
\end{equation}
where we have also written $w=ae^{iq}$.  In view of \eqref{eq:dA}, and
the periodicity of $r$, the first integral term on the RHS in
\eqref{eq:dr0} vanishes for area-preserving perturbations.  The last
integral in \eqref{eq:dr0} corresponds to the second to last in
\eqref{eq:dr} and the last integral in \eqref{eq:dr} vanishes (this
follows from the explicit formula for principal-value integrals that
we give below).  We note that relation \eqref{eq:dr0} is equivalent to
the perturbation equation obtained with the Lagrangian approach by
Baker in his analysis of the stability of the Rankine vortex
\cite{b90}. As was done in earlier studies \cite{k80,l32,b67,s92,b90},
we will adopt the ``normal mode'' approach and suppose that
\begin{equation}
r(\tau,p) = \rho(\tau) \, e^{ikp} + \overline{\rho}(\tau) \, e^{-ikp},
\label{eq:nmode}
\end{equation}
where $\rho(t) \in \CC$ and $k \in \ZZ$. As will be demonstrated
below, the normal mode approach is justified, because equation
\eqref{eq:dr0} does not lead to mode coupling. Using ansatz
\eqref{eq:nmode}, the second integral in \eqref{eq:dr0} becomes
\begin{equation}
\begin{aligned}
& \int_0^{2\pi}\frac{(e^{iq} + e^{is}) r(q)}{e^{is} - e^{iq}}\,dq = \\
& \rho(t) \left[ 
\int_0^{2\pi}\frac{e^{i(k+1)q}}{e^{ip} - e^{iq}}\,dq +
e^{ip} \int_0^{2\pi}\frac{e^{ikq}}{e^{ip} - e^{iq}}\,dq \right] + \\
& \overline{\rho}(t) \left[  
\int_0^{2\pi}\frac{e^{i(1-k)q}}{e^{ip} - e^{iq}}\,dq +
e^{ip} \int_0^{2\pi}\frac{e^{-ikq}}{e^{ip} - e^{iq}}\,dq \right].
\end{aligned}
\label{eq:pvi}
\end{equation}
Each of the integrals on the RHS in \eqref{eq:pvi} can be written in
the form $\int_0^{2\pi}\frac{e^{iMq}}{e^{is} - e^{iq}}\,dq$ for some
$M \in \ZZ$. This principal-value integral can be evaluated
analytically using contour integration (see Appendix \ref{sec:pvint})
and the result is
\begin{equation}
\int_0^{2\pi}\frac{e^{iMq}}{e^{ip} - e^{iq}}\,dq \quad = \quad 
\left\{ 
\begin{alignedat}{2}
-& \pi \, e^{i \, (M-1) \, p}, \qquad && M \ge 1, \\
 & \pi \, e^{i \, (M-1) \, p}, && M < 1. \\
\end{alignedat}\right.
\label{eq:pvM}
\end{equation}
Inserting ansatz \eqref{eq:nmode} into perturbation equation
\eqref{eq:dr0} and then using \eqref{eq:pvM} to reduce \eqref{eq:pvi}
we obtain an equation in which all terms are proportional either
to $e^{iks}$, or to $e^{-iks}$. The fact that there are
no other terms justifies a posteriori the normal mode approach
\eqref{eq:nmode}. Finally, isolating the terms proportional to
$e^{iks}$ and $e^{-iks}$ we obtain
\begin{equation}
\frac{d}{dt} \rho(\tau)  = - \frac{i \omega}{2} k\, \rho(\tau) + \frac{i \omega}{2} \, \rho(\tau)
\label{eq:rho} 
\end{equation}
and its complex conjugate copy, so that the solution is $\rho(\tau) =
\rho_0 \, e^{\frac{i \omega}{2} (1-k)\, \tau}$, where $\rho_0 \in \CC$
is the amplitude of the initial perturbation. From \eqref{eq:nmode} we
obtain the following expression for the evolution of the perturbation
\begin{equation}
r(\tau,p) = \rho_0 \, e^{i\left[\frac{\omega}{2} (1-k)\, \tau + k\, p\right]} + 
\overline{\rho}_0 \, e^{-i\left[\frac{\omega}{2} (1-k)\, \tau + k\, p\right]}
\label{eq:rt}
\end{equation}
in which the phase velocity $v_p = \frac{\omega}{2}\frac{k-1}{k}$
agrees with Kelvin's solution \cite{k80,l32,b67,s92}. We add that the
problem analyzed by Lamb \cite{l32} is in fact formulated in a bit
different way, because he assumed the perturbation in the form of a
{\em potential} modification of the velocity field, so that the shape
of the region $A$ where $\omega \neq 0$ remains unchanged \cite{l32}.
It is therefore interesting that the solutions obtained in these two
problems are identical, {despite subtle differences in the
  underlying assumptions}.

\section{Stability of Elliptic Vortex}
\label{sec:elliptic}

In this Section we demonstrate how the well-known results concerning
the linear stability analysis of Kirchhoff's elliptic vortex
\cite{k76} obtained initially by Love \cite{l93} can be analytically
deduced from general perturbation equation \eqref{eq:dr}. Let the
vortex boundary be described as
\begin{equation}
z(p) = a\, \cos(p) + i \, b \, \sin(p), \qquad p \in [0,2\pi],
\label{eq:zq}
\end{equation}
where $a$ and $b$ are, respectively, the major and minor axes of the
ellipse, and $p$ coincides with one of the coordinates in the elliptic
coordinate system. The key result is that the flow becomes linearly
unstable when the ellipse aspect ratio $c = a/b > 3$. As the aspect
ratio increases, consecutive eigenmodes are destabilized. They are
given in the form \cite{l93,mr08,ghs04}
\begin{equation}
r_k(p) = \Big| \frac{dz}{dp} \Big|^{-1} \left[\cos(k \, p) - 
\sqrt{\bigg|\frac{\mu_m^-}{\mu_m^+}\bigg|}\,\sin(k \, p)\right], \qquad k=3,\dots,
\label{eq:rk}
\end{equation}
where 
\begin{equation}
\mu_m^+ = \frac{\omega}{2} \left[ \left(\frac{ c - 1 }{ c + 1 }\right)^k + 
\left( \frac{2 \, k \, c}{ ( c + 1 )^2} - 1 \right) \right], \quad
\mu_m^- = \frac{\omega}{2} \left[ \left(\frac{ c - 1 }{ c + 1 }\right)^k -
\left( \frac{2 \, k \, c}{ ( c + 1 )^2} - 1 \right) \right].
\label{eq:mu}
\end{equation}
As can be verified, representation \eqref{eq:rk} satisfies the
  area-preservation condition \eqref{eq:dA}.  The associated
eigenvalues are
\begin{equation}
\lambda_k = \sqrt{ - \mu_m^+ \, \mu_m^- } = 
\frac{\omega}{2}  \left[ \left( \frac{2 \, k \, c}{ ( c + 1 )^2} - 1 \right)^2  
- \left(\frac{ c - 1 }{ c + 1 }\right)^{2k} \right],
\label{eq:lambda}
\end{equation}
so that eigenmodes \eqref{eq:rk} grow proportionally to $e^{-i
  \lambda_k t}$. We note that only modes with $k\ge 3$ can become
unstable, and to every such mode there corresponds a decaying mode
with $\lambda_k = - \sqrt{ - \mu_m^+ \, \mu_m^- }$. These results are
deduced from perturbation equation \eqref{eq:dr} by plugging $r(t,p) =
e^{-i \lambda_k t} \, r_k(p)$ into this equation, where $r_k(p)$ is
given in \eqref{eq:rk}, and then performing the following steps
\begin{enumerate}
\item substitute representation \eqref{eq:zq} for $z=z(p)$ and
    $w=w(q)$ in \eqref{eq:dr} and convert the contour integrals to
  definite ones with $q \in [0,2\pi]$ as the integration
    variable,

\item since $p$ and $q$ appear only as arguments of trigonometric
  functions, use the following identities ($u \in \CC$)
\begin{equation}
\cos(k\,q) = \frac{u^k+u^{-k}}{2}, \quad \sin(k\,q) = \frac{u^k-u^{-k}}{2i}
\label{eq:coskq}
\end{equation}
and likewise for $\cos(k\,p)$ and $\sin(k\,p)$ using $v\in\CC$,
to transform the resulting expressions to contour integrals over the
unit circle $|u|=1$ with $du = i \, u \, dp$,

\item perform partial fraction decomposition of thus obtained
  integrand {expressions} and evaluate the integrals using the
  residue theorem for the poles $u_0$ inside the unit circle
  ($|u_0|<1$) and formula \eqref{eq:pvi} for the poles $u_0$ on the
  unit circle ($|u_0|=1$).
\end{enumerate}
For any given value of $k$, operations described in steps 1.--3. above
can be performed symbolically using a symbolic algebra package such as
{\tt Maple} (the corresponding code is available on-line as
  supplementary material). Combining all resulting terms and noting
that for the elliptic vortex $V_t(p) = \Re[V t] = \frac{\omega}{a+b}
\frac{a^2 \, b^2 \, (a\,\sin(p)^2 + b\, \cos(p)^2) }{\sqrt{a^4 \,
    b^2\,\sin(p)^2 + b^4\,a^2\,\cos(p)^2)}}$, cf.~\eqref{eq:V}, and
$V_{0}(p) = - i \omega \frac{ab}{(a+b)^2} \,\overline{z}(p)$,
cf.~\eqref{eq:V1}, we recover expression \eqref{eq:lambda} for the
eigenvalue. We add that in the special case when $a=b$ results
of Section \ref{sec:circular} are obtained, except that in this limit
the velocity characterizing the rotation of the coordinate system is
$V_{0t} = \frac{\omega \, a}{4}$.

\section{Accurate Numerical Solution of Stability Equations}
\label{sec:numer}

In this Section we describe a spectrally-accurate numerical technique
which allows us to solve perturbation equation \eqref{eq:dr}
numerically for vortices of arbitrary smooth shape. A key feature of
this approach is that the singular parts of the integrals in
\eqref{eq:dr} are in fact evaluated analytically. {We remark that
  such an approach was also used in other stability analyses of 2D
  vortex flows, e.g., in \cite{d95}.}  Our method is validated by
comparing the results obtained numerically using different resolutions
for the elliptic vortex with the analytical results derived by Love
\cite{l93}, and discussed in Section \ref{sec:elliptic}. First, we
assume that the vortex boundary $\partial A$ admits a smooth periodic
and otherwise arbitrary parametrization $z = z(p)$, $p \in [0,2\pi]$.
In order to simplify the notation, we rewrite perturbation equation
\eqref{eq:dr} as
\begin{equation}
\begin{aligned}
\Dpartial{r(p)}{\tau} =  & - V_{1t}(p)\,  \Big| \frac{dz}{dp}(p) \Big|^{-1} \, \Dpartial{r(p)}{p} \\
& +  \Re\left[ (\L_1 \, r)(p) + (\L_2 \, r)(p) + (\L_3 \, r)(p) + (\L_4 \, r)(p) \right],
\qquad p \in [0,2\pi],
\end{aligned}
\label{eq:dr2}
\end{equation}
where the dependence of the perturbation $r$ on time $\tau$ was
suppressed for brevity, and $\L_1 \, r \triangleq \Re\left[n \, U'(z)
  + rn \left( \frac{dU}{dz} - \frac{dV_0}{dz}\right)\right]$, whereas
$\L_2 \,r$, $\L_3 \,r$ and $\L_4 \,r$ represent, respectively, the
three integral operators on the RHS of \eqref{eq:dr}.  Quantities
appearing in equation \eqref{eq:dr2} which are related to the contour
geometry, i.e., $\big| \frac{dz}{dp} \big|^{-1}$, $t$, $n$ and
$\kappa$, can be evaluated using spectral differentiation. Equation
\eqref{eq:dr2} is discretized by using the following ansatz for the
perturbation
\begin{equation}
r(\tau,p) \approx r^N(\tau,p) = e^{i \, \lambda \, \tau} \,
\Big| \frac{dz}{dp} \Big|^{-1} \, 
\left[ \sum_{k=0}^{N/2} \alpha_k \cos(k\, p) + \sum_{k=1}^{N/2-1} \beta_k \sin(k\, p) \right]
\label{eq:rn}
\end{equation}
depending on $N$ real-valued coefficients $\alpha_k$ and $\beta_k$ and
where $\lambda \in \CC$ is the growth rate we seek to determine.  The
factor $\big| \frac{dz}{dp} \big|^{-1}$ in \eqref{eq:rn} ensures that
area-preservation condition \eqref{eq:dA} is satisfied by all terms in
the series, except for the constant term $\alpha_0 \cos(0 \, p)$ which
must be introduced in order to ensure well-posedness of the
collocation problem (it will be, however, subtracted via projection
from the resulting eigenvectors). After using the ansatz \eqref{eq:rn},
the independent variable $p$ is discretized using $N$ equispaced
collocation points $\{ p_j \}_{j=1}^N$ ($N$ is an even number).
Denoting $\y =
[\alpha_0,\dots,\alpha_{N/2},\beta_1,\dots,\beta_{N/2-1}]^T \in
\RR^N$, this discretization leads to the following eigenvalue problem
\begin{equation}
\lambda \, \y = - i \, \A^{-1} (\B + \C) \y,
\label{eq:eval}
\end{equation}
where 
\begin{align}
[\A]_{jl} & = r_l(p_j), \qquad j,l=1,\dots,N, \label{eq:A} \\
\textrm{in which} \ r_l(p) & = \left\{ 
\begin{alignedat}{2}
& \cos((l-1)\, p), & \quad & l=1,\dots,\frac{N}{2}+1, \\
& \sin\left[\left(l-\frac{N}{2}-1\right)\, p \right], && l=\frac{N}{2}+2,\dots,N
\end{alignedat}\right.,
\label{eq:rl}
\end{align}
is the (invertible) collocation matrix,
\begin{equation}
[\B]_{jl} = - V_{1t}(p_j) \, \Big| \frac{dz}{dp}(p_j) \Big|^{-1} \, 
\sum_{m=1}^N \A_{jm} \, \D_{ml}, \qquad  j,l=1,\dots,N
\label{eq:B}
\end{equation}
corresponds to the first term on the RHS in \eqref{eq:dr2} with $\D$
denoting the spectral differentiation matrix associated with ansatz
\eqref{eq:rn}, and
\begin{equation}
[\C]_{jl} = \Re\left[ (\L_1 \, r_l)(p_j) + (\L_2 \, r_l)(p_j) + (\L_3 \, r_l)(p_j) + (\L_4 \, r_l)(p_j) \right], \qquad  j,l=1,\dots,N
\label{eq:C}
\end{equation}
in which the first term on the RHS is obtained via straightforward
evaluation whereas the calculation of the integral expressions
$\L_2\,r_l$, $\L_3\,r_l$, and $\L_4\,r_l$ is described below.

All integrals with bounded integrands are evaluated using the
trapezoidal rule, which on a periodic domain is a spectrally-accurate
quadrature. Since $p$ and $q$ are now the independent variables, in
this Section these characters will be used as subscripts to indicate
where a given variable is evaluated. The integrand expressions in
$V_{1t}$, cf.~\eqref{eq:V}, and in $\L_2 \, r \triangleq -
\frac{\omega \, n_p}{4 \pi} \int_0^{2\pi} \frac{\overline{z} -
  \overline{w_q}}{z - w_q} \left(i \, \big| \frac{dz}{dq} \big|^{-1}
  \, \frac{dr}{dq} - \kappa\,r_q\right)\,t_q\,\big| \frac{dz}{dq}
\big|\, dq$ have removable singularities and are therefore bounded
\cite{s92}. As regards the integral operators $\L_3$ and $\L_4$ which
are defined in the principal-value sense, we use a standard approach
(see, e.g., \cite{h95}) in which they are split into principal-value
integrals corresponding to a circle, which can be evaluated
analytically using formula \eqref{eq:pvM}, and regular ``corrections''
which can be approximated numerically with the spectral accuracy using
the trapezoidal rule. In order to be able to use \eqref{eq:pvM}, the
basis functions in ansatz \eqref{eq:rn} are expressed in terms of the
complex exponentials, so that
\begin{align*}
\L_3 \, \left(\Big| \frac{dz}{dq} \Big|^{-1} \cos(k \, q)\right) 
& = \phantom{=} \frac{1}{2}\left[ \L_3 \,
  \left(\Big| \frac{dz}{dq} \Big|^{-1} e^{ikq}\right) + \L_3 \,
  \left(\Big| \frac{dz}{dq} \Big|^{-1} e^{-ikq}\right)\right], \\
\L_3 \, \left(\Big| \frac{dz}{dq} \Big|^{-1} \sin(k \, q)\right) 
& = -\frac{i}{2}\left[ \L_3 \, \left(\Big| \frac{dz}{dq} \Big|^{-1}
    e^{ikq}\right) - \L_3 \, \left(\Big| \frac{dz}{dq} \Big|^{-1}
    e^{-ikq}\right)\right],
\end{align*}
and analogously for $\L_4\, r$. Then, we have
\begin{equation}
\begin{aligned}
\L_4 \, & \left(\Big| \frac{dz}{dq} \Big|^{-1} e^{imq} \right)(p) 
=  - \frac{\omega \, n_p}{4  \pi} \Big| \frac{dz}{dq}(p) \Big|^{-1} e^{imp}  
\int_0^{2\pi} \frac{ \frac{\overline{z} - \overline{w_q}}{z - w_q} \, n_p - \overline{n}_p}{z_p - w_q} \, \frac{dw}{dq} \, dq \\
& = - \frac{\omega \, n_p}{4  \pi} \Big| \frac{dz}{dq}(p) \Big|^{-1} e^{imp} \left\{
\int_0^{2\pi} \left[ \frac{\frac{dw}{dq}}{z_p - w_q} - \frac{i\, e^{iq}}{e^{ip} - e^{iq}} \right] \, Q(p,q)\, dq 
+ i \int_0^{2\pi} \frac{e^{iq}}{e^{ip} - e^{iq}} \, Q(p,q)\, dq\right\}, 
\end{aligned}
\label{eq:L4}
\end{equation}
where $Q(p,q) \triangleq \frac{\overline{z} - \overline{w_q}}{z -
  w_q}\, n_p - \overline{n}_p$. Expanding this expression in a Fourier
series with respect to $q$, i.e., $Q(p,q) = \sum_{k=-N/2}^{N/2}
\hat{Q}_k(p) \, e^{ikq}$, the second integral on the RHS in
\eqref{eq:L4} can be evaluated as follows with the help of formula
\eqref{eq:pvM}
\begin{equation}
\begin{aligned}
i \int_0^{2\pi} \frac{e^{iq}}{e^{ip} - e^{iq}} \, Q(p,q)\, dq 
& = i \sum_{k=-N/2}^{N/2} \hat{Q}_k(p) \, \int_0^{2\pi} \frac{e^{i(k+1)q}}{e^{ip} - e^{iq}} \,dq \\
& = i \pi  \left[ \sum_{k=-N/2}^{-1} \hat{Q}_k(p) \, e^{ikq} - \sum_{k=0}^{N/2} \hat{Q}_k(p) \, e^{ikq}\right].
\end{aligned}
\label{eq:L4b}
\end{equation}

As regards the remaining integral operator, we have
\begin{equation}
\begin{aligned}
\L_3 \,  \left(\Big| \frac{dz}{dq} \Big|^{-1} e^{imq} \right)(p) 
= & \frac{\omega \, n_p}{4  \pi}   
\int_0^{2\pi} \frac{ \big| \frac{dz}{dq}(q) \big|^{-1} e^{imq}\, 
\frac{\overline{z} - \overline{w_q}}{z - w_q} \, n_q 
- \overline{n}_q}{z_p - w_q} \, \frac{dw}{dq} \, dq \\
= & \frac{\omega \, n_p}{4  \pi} \Bigg\{
\int_0^{2\pi} \left[ \frac{\frac{dw}{dq}}{z_p - w_q} - \frac{i\, e^{iq}}{e^{ip} - e^{iq}} \right] \,
e^{imq} \, \Big| \frac{dz}{dq}(q) \Big|^{-1} \, S(p,q)\, dq  \\
& + i\, \int_0^{2\pi} \frac{e^{iq}}{e^{ip} - e^{iq}} e^{imq} \, \Big| \frac{dz}{dq}(q) \Big|^{-1} \, S(p,q)\, dq\Bigg\}, 
\end{aligned}
\label{eq:L3}
\end{equation}
where $S(p,q) \triangleq \frac{\overline{z} - \overline{w_q}}{z - w_q}\,
n_q - \overline{n}_q$. Expanding this expression multiplied $\big|
\frac{dz}{dq}(q) \big|^{-1}$ in a Fourier series with respect to $q$,
i.e., $\big| \frac{dz}{dq}(q) \big|^{-1} \,S(p,q) = \sum_{k=-N/2}^{N/2}
\left(\widehat{\big| \frac{dz}{dq} \big|^{-1} S(p)}\right)_k \,
e^{ikq}$, the second integral on the RHS in \eqref{eq:L3} can be
evaluated as follows with the help of formula \eqref{eq:pvM}
\begin{equation}
\begin{aligned}
i \int_0^{2\pi} \frac{e^{iq}}{e^{ip} - e^{iq}} e^{imq} \, & \Big| \frac{dz}{dq}(q) \Big|^{-1} \, S(p,q)\, dq
= i \sum_{k=-N/2}^{N/2}  \left(\widehat{\Big| \frac{dz}{dq} \Big|^{-1} S(p)}\right)_k\, 
\int_0^{2\pi} \frac{e^{i(k+m+1)q}}{e^{ip} - e^{iq}} \,dq \\
& = i \pi  \left[ \sum_{k=-N/2}^{-m-1} \left(\widehat{\Big| \frac{dz}{dq} \Big|^{-1} S(p)}\right)_k \, e^{ikq} 
- \sum_{k=-m}^{N/2} \left(\widehat{\Big| \frac{dz}{dq} \Big|^{-1} S(p)}\right)_k \, e^{ikq}\right].
\end{aligned}
\label{eq:L3b}
\end{equation}
The first integrals on the RHS in \eqref{eq:L4} and \eqref{eq:L3} are
regular and can be evaluated numerically (they also vanish in the
circular case). Expressions $\L_3 \, \left(\big| \frac{dz}{dq}
  \big|^{-1} e^{-imq} \right)$ and $\L_4 \, \left(\big| \frac{dz}{dq}
  \big|^{-1} e^{-imq} \right)$ are computed in an analogous manner.
After collecting expressions \eqref{eq:A}--\eqref{eq:L3b}, the
algebraic eigenvalue problem \eqref{eq:eval} can be solved using
standard tools.

\begin{figure}[t]
\begin{center}
\includegraphics[width=0.9\textwidth]{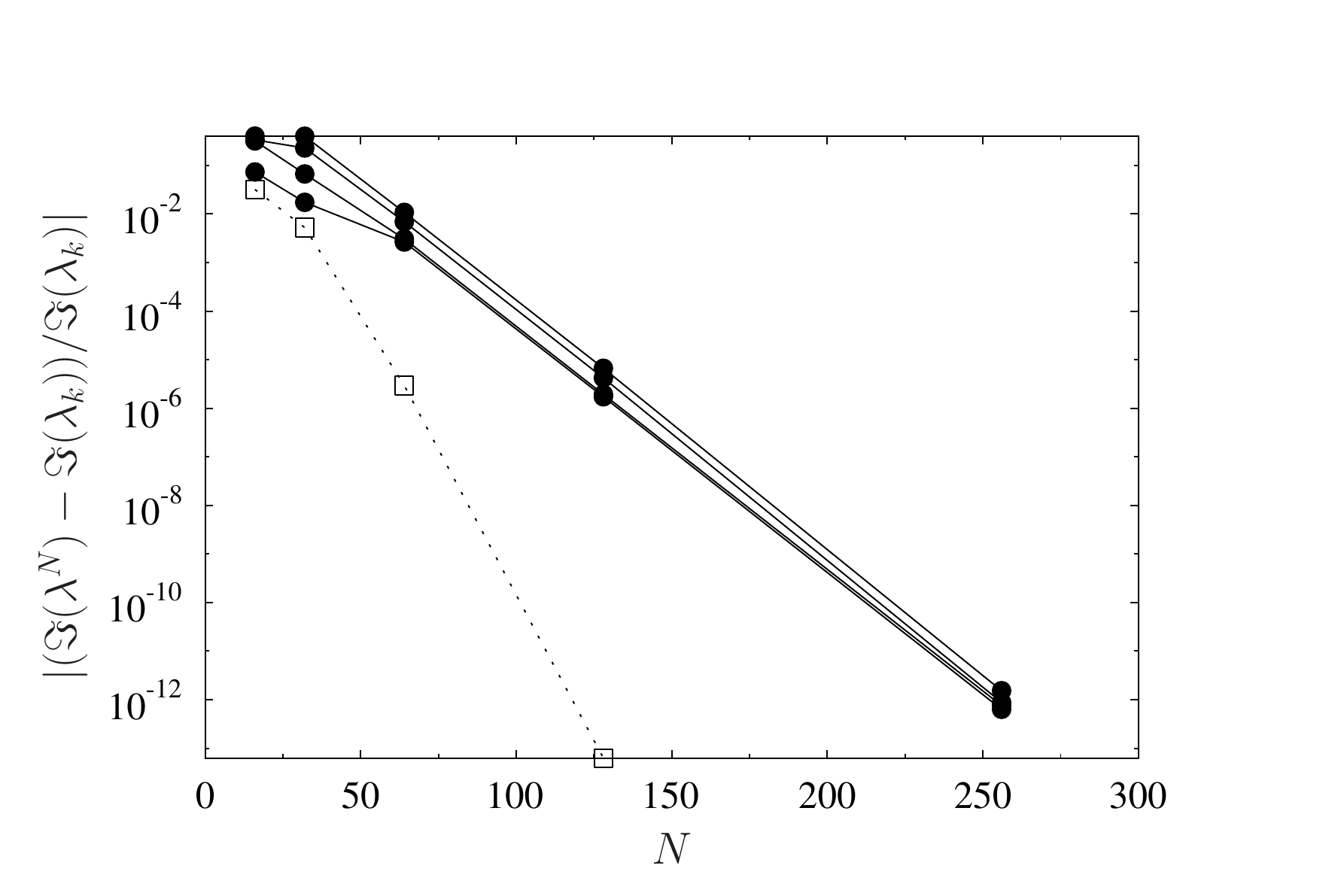}
\caption{Relative errors in the imaginary parts of the numerically
  computed eigenvalues $\lambda^N$ with respect to the analytical
  expressions for $\lambda_k$ given in equation \eqref{eq:lambda},
  cf.~\cite{l93}, as a function of the resolution $N$ for vortices
  with (open symbols) aspect ratio $a/b=4$ and (solid symbols) aspect
  ratio $a/b=8$. {For the aspect ratio $a/b=4$ the eigenmode
    wavenumber corresponding to the eigenvalue is $k=3$, whereas for
    the aspect ratio $a/b=8$ the wavenumbers are $k=3,4,5,6$,
    cf.~\eqref{eq:rn}, with larger values of $k$ resulting in larger
    errors.}}
\label{fig:evals_Love}
\end{center}
\end{figure}

We close this Section with a numerical validation of the proposed
method. As the test problem we used the case of the elliptic vortex,
also studied analytically in Section \ref{sec:elliptic}. We compute
the eigenvalues by solving problem \eqref{eq:eval} using different
numerical resolutions $N$ and compare the results against the
closed-form expression \eqref{eq:lambda} obtained by Love \cite{l93},
see also \cite{ghs04,mr08}. More specifically, we analyze the
imaginary parts $\Im(\lambda)$ of the eigenvalues responsible for the
instability. In Figure \ref{fig:evals_Love} we show the relative
errors between the eigenvalues $\Im(\lambda^N)$ computed numerically
with resolution $N$ and $\Im(\lambda_k)$ obtained analytically by
Love, cf.~\eqref{eq:lambda}, for two different aspects ratios of the
elliptic vortex (for the aspect ratio equal to 8 there are in fact 4
distinct unstable eigenvalues).  In Figure \ref{fig:evals_Love} we
note that this error decays exponentially fast dropping to the machine
precision level already for modest resolutions $N$, thereby confirming
the spectral accuracy of the proposed method. We add that, as
expected, the resolutions required to achieve a given accuracy
increase with the aspect ratio of the vortex.

\section{Conclusions}
\label{sec:final}

While perturbation equations similar to \eqref{eq:dr} have already
been used for the linear stability analysis of 2D vortex patches
\cite{d85,dl91,d95}, to the best of our knowledge, the present study
{offers} a first {complete} derivation of this approach. In
particular, it relies on methods of the shape calculus which are a
general and mathematically consistent way of dealing with the
free-boundary aspect of the problem. We add that, in the context of
vortex dynamics, such techniques have already been used to study
continuation of families of solutions \cite{gipz09a} and vortex
control problems \cite{p09a}. The proposed numerical approach is
demonstrated to be spectrally-accurate reducing all singular integrals
to closed form \eqref{eq:pvi}. The two validation tests presented
offer ``the proof of the concept'' for this approach.

Generalization of our method to problems involving several vortices is
straightforward, and requires that an equation of the form
\eqref{eq:dr} be written for each individual vortex with the
interaction between the vortices captured by the field $U(z)$ (which
vanishes identically in the single-vortex examples considered in
Sections \ref{sec:circular} and \ref{sec:elliptic}). This description
will be simplified by symmetries of the vortex configuration. One
aspect of this problem which does not seem to have received much
attention in the literature is the possibility for subcritical
disturbance amplification due to non-normality of the underlying
stability operator. While these questions have already been
investigated for viscous vortices with unbounded vorticity support
(e.g., \cite{ph06,msb12}), to the best of our knowledge, there are no
results concerning free-boundary problems of the type
\eqref{eq:euler}--\eqref{eq:F}. Another interesting and far less
researched application area for our approach is the stability analysis
of 3D axisymmetric vortex flows, with and without swirl, with compact
vorticity support \cite{efm08}. All these problems are left to future
research.

As regards limitations of the proposed approach, we remark that shape
calculus in its standard formulation \cite{dz04} is applicable
only to problems posed on smooth manifolds, hence our method
would need to be modified, so that it can be applied to contours with
singularities, such as, e.g., corners (flows with such features
typically arise as terminal members of families of steady solutions
\cite{gipz09a}).  Some relevant ideas are already mentioned in
\cite{em95}.  Likewise, analysis of stability with respect to
perturbations leading to topology changes requires the use of
different methods.

%\ack{
\section*{Acknowledgments}
The authors are grateful to Prof.~L.~Rossi for insightful comments
about Love's stability analysis of the elliptic vortex \cite{l93}.  BP
was partially supported through an NSERC (Canada) Discovery Grant.
%}

\appendix

\section{Evaluation of the Integral $\int_0^{2\pi}\frac{e^{iMq}}{e^{is} - e^{iq}}\,dq$}
\label{sec:pvint}

The calculations presented in Section \ref{sec:circular},
\ref{sec:elliptic} and \ref{sec:numer} required the values of the
integrals
\begin{equation}
I=\int_{0}^{2\pi}\frac{e^{iM}}{e^{ip}-e^{iq}} \, dq
=e^{-ip}\int_{0}^{2\pi}\frac{e^{iM}}{1-e^{i(q-p)}}\, dq
\label{eq:I}
\end{equation}
for integer $M$. If $M>0$, let us consider
\begin{equation*}
e^{-ip}\int_{C}\frac{e^{iM}}{1-e^{i(q-p)}}\,dq
\end{equation*}
in which $C$ is now the contour connecting the points $0$, $2\pi$,
$2\pi+Yi$, $Yi$, and $0$ with a small semi-circular indentation into
the upper half-plane above $p$, where $p$ $\in (0,2\pi)$. This contour
integral vanishes by Cauchy's Theorem. The integrals over the left and
right lateral sides of $C$ cancel by periodicity. As regards the
  top segment of contour $C$, writing $q=x+iY$, we have
\begin{equation*}
e^{iMq}=e^{iMx}e^{-MY} \quad \textrm{and} \quad  e^{i(q-p)}=e^{i(x-p)}e^{-Y},
\end{equation*}
so for $Y \rightarrow \infty$ the integral over that segment
goes to zero. There remains to take the limit of the integral over the
bottom segment of the contour $C$ as the radius of the semi-circular
indentation vanishes. This is the original principal-value integral
\eqref{eq:I} augmented by the contribution from the indentation, which
can be evaluated using the Cauchy integral formula. To this end, we
write the integrand as
\begin{equation*}
\frac{e^{iMq}(q-p)}{1-e^{i(q-p)}}\frac{1}{q-p}
\end{equation*}
and find, using L'H\^opital's rule,
\begin{equation*}
\lim_{q \rightarrow p}\frac{e^{iMq}(q-p)}{1-e^{i(q-p)}} =  ie^{i(M-1)p}.
\end{equation*}
Thus, the limit of the integral over the indentation is this value
multiplied by $-\pi i$, and, finally, we obtain
\begin{equation}
I = \pi i (i e^{i(M-1)p}) = -\pi e^{i(M-1) p}.
\label{eq:pvi1}
\end{equation}
When $M<1$, instead of contour $C$, we use its reflection into the
lower half-plane. Then, since $Y$ is negative, the
integral over the bottom segment goes to zero as $Y \rightarrow
-\infty$ and, since the small semi circle is positively oriented, 
\begin{equation}
I = - \pi i (i e^{i(M-1)p}) = \pi e^{i(M-1)p}.
\label{eq:pvi2}
\end{equation}
Formulas \eqref{eq:pvi1} and \eqref{eq:pvi2} are equivalent to
  \eqref{eq:pvM}.


\begin{thebibliography}{100}
%\bibitem[Kelvin (1880)]{k80}
%Lord Kelvin, 1880 ``Vibrations of a columnar vortex'', {\em Phil.~Mag}
%{\bf{10}}, 155--168.

%\bibitem[Lamb (1932)]{l32}
%H.~Lamb, {\em Hydrodynamics}, Cambridge University press, (1932).

%\bibitem[Batchelor (1967){b67}
%G.~K.~Batchelor, {\em An introduction to fluid mechanics}, Cambridge
%University Press, (1967).

%\bibitem[Saffman (1992)]{s92}
%P.~G.~Saffman, {\em Vortex Dynamics}, Cambridge University Press, (1992).

\bibitem{k80}
Lord Kelvin, ``Vibrations of a columnar vortex'', {\em Phil.~Mag}
{\bf{10}}, 155--168, (1880).

\bibitem{l32}
H.~Lamb, {\em Hydrodynamics}, Cambridge University press, (1932).

\bibitem{b67}
G.~K.~Batchelor, {\em An introduction to fluid mechanics}, Cambridge
University Press, (1967).

\bibitem{s92}
P.~G.~Saffman, {\em Vortex Dynamics}, Cambridge University Press, (1992).

\bibitem{b90}
G.~R.~Baker, ``A Study of the Numerical Stability of the Method of
Contour Dynamics'', {\em Phil. Trans. Roy. Soc.} {\bf{333}}, 391--400,
(1990).

\bibitem{k76} G.~Kirchhoff, {\em Vorlesungen uber Mathematische
    Physik: Mechanik}, Teubner, (1876).

\bibitem{l93} A.~E.~H. Love, ``On the Stability of certain Vortex
  Motions'', {\em Proc. London Math. Soc.}, {\bf{s1--25}}, 18-43,
  (1893).

\bibitem{ms71}
D.~W.~Moore and P.~G.~Saffman, ``Structure of a line vortex in an
imposed strain'', in Olsen, Goldburg and Rogers (Eds.) {\em Aircraft
wake turbulence}, Plenum, 339--354, (1971).

\bibitem{mr08} T.~B.~Mitchell and L.~F.~Rossi, ``The evolution of
  Kirchhoff elliptic vortices'', {\em Phys. Fluids} {\bf{20}}, 054103
  (2008).

\bibitem{ghs04} Y.~Guo, Ch.~Hallstrom, and D.~Spirn, ``Dynamics Near
  an Unstable Kirchhoff Ellipse'' {\em Commun. Math. Phys.}
  {\bf{245}}, 297--354, (2004).

\bibitem{d85}
D.~G.~Dritschel, ``The stability and energetics of corotating uniform
vortices'', {\em J. Fluid Mech.} {\bf{157}}, 95--134, (1985).

\bibitem{d90} D.~G.~Dritschel, ``The stability of elliptical vortices
  in an external straining flow'', {\em J. Fluid Mech.} {\bf{210}},
  223--261, (1990).

\bibitem{d95}
D.~G.~Dritschel, ``A general theory for two--dimensional vortex
interactions'', {\em J. Fluid Mech.} {\bf{293}}, 269--303, (1995).

\bibitem{dl91}
D.~G.~Dritschel and B.~Legras, ``The elliptical models of
two--dimensional vortex dynamics. II: Disturbance equations'', {\em
Phys. Fluids A} {\bf{3}}, 855--869, (1991).

\bibitem{d92} M.~R.~Dhanak, ``Stability of a Regular Polygon of Finite
  Vortices'', {\em J. Fluid Mech.} {\bf{234}}, 297--316, (1992).

\bibitem{woz84}
H.~M.~Wu, E.~A.~Overman II, and N.~J.~Zabusky, ``Steady--State
Solutions of the Euler Equations in Two Dimensions: Rotating and
Translating V--States and Limiting Cases. I. Numerical Algorithms and
Results'', {\em J.~Comp.~Phys.} {\bf{53}}, 42--71, (1984).

\bibitem{k87}
J.~R.~Kamm, ``Shape and stability of two--dimensional vortex
regions'', Ph.D.~Thesis, Caltech, (1987).

\bibitem{efm05}
A.~Elcrat, B.~Fornberg B and K.~Miller, ``Stability of vortices in
equilibrium with a cylinder'', {\it J. Fluid Mech.} {\bf 544}, 53--68,
(2005).

\bibitem{b82}
J.~Burbea, ``On Patches of Uniform Vorticity in a Plane of
Irrotational Flow'', {\em Arch. Rat. Mech. Anal.} {\bf{77}}, 349--358,
(1982).

\bibitem{bl82}
J.~Burbea and M.~Landau, ``The Kelvin Waves in Vortex Dynamics and
Their Stability'', {\em J. Comp. Phys} {\bf{45}}, 127--156, (1982).

\bibitem{t85}
B.~Turkington, ``Corotating Steady Vortex Flows with $N$--Fold
Symmetry'', {\em Nonlinear Analysis, Theory, Methods \& Applications}
{\bf{9}}, 351--369, (1985).

\bibitem{t92}
Y.~Tang, ``Nonlinear Stability of Rankine's Vortex'', {\em
Int. J. Non--Linear Mech.} {\bf{27}}, 669--673, (1992).

\bibitem{w86}
Y.-H.~Wan, ``The Stability of Rotating Vortex Patches'', {\em
Commun. Math. Phys.} {\bf{107}}, 1--20, (1986).

\bibitem{w98}
Y.-H.~Wan, ``Bifurcations at Kirchhoff elliptic vortex with
eccentricity $2\sqrt{2}/3$'', {\em Dynamics and Stability of Systems''}
{\bf{13}}, 281--297, (1998).

\bibitem{ss80}
P.~G.~Saffman and R.~Szeto, ``Equilibrium shapes of a pair of equal
uniform vortices'', {\em Phys Fluids} {\bf{23}}, 2339--2342, (1980).

\bibitem{d88} D.~G.~Dritschel, ``Nonlinear stability bounds for
  inviscid, two-dimensional, parallel or circular flows with monotonic
  vorticity, and the analogous three- dimensional quasi-geostrophic
  flows'', {\em J. Fluid Mech.} {\bf{191}}, 575--581, (1988).


\bibitem{cw03}
C.~Cerrtelli and C.~H.~K.~Williamson, ``A new family of inform
vortices related to vortex configurations before merging'', {\it
J. Fluid Mech.} {\bf 493}, 219--229, (2003).

\bibitem{fm08} Y.~Fukumoto \& H.~K.~Moffatt, ``Kinematic variational
  principle for motion of vortex rings'', {\em Physica D} {\bf{237}},
  2210--2217, (2008).

\bibitem{fw10a} P.~Luzzatto-Fegiz and C.~H.~K.~Williamson, ``Stability
  of Conservative Flows and New Steady-Fluid Solutions from
  Bifurcation Diagrams Exploiting Variational Argument'', {\em Phys.
    Rev. Lett.} {\bf{104}}, 044504, (2010).

\bibitem{fw10b} P.~Luzzatto-Fegiz and C.~H.~K.~Williamson, ``Stability
  of elliptical vortices from imperfect-velocity-impulse diagrams''.
  {\em Theor. Comput. Fluid Dyn.} {\bf{24}}, 181--188, (2010).

\bibitem{fw11a} P.~Luzzatto-Fegiz and C.~H.~K.~Williamson, ``An
  accurate and efficient method for computing uniform vortices'', {\em
    J. Comput. Phys.} {\bf{230}}, 6495--6511 (2011).

\bibitem{fw11b} P.~Luzzatto-Fegiz and C.~H.~K.~Williamson,''Resonant
  instability in two-dimensional vortex arrays'', {\em Proc. R. Soc.
    A} {\bf{467}}, 1164--1185, (2011).

\bibitem{fw12a} P.~Luzzatto-Fegiz and C.~H.~K.~Williamson, ``Structure
  and stability of the finite-area K\'arm\'an street'', {\em Phys.
    Fluids} {\bf{24}}, 066602, (2012).

\bibitem{fw12b} P.~Luzzatto-Fegiz and C.~H.~K.~Williamson,
  ``Determining the stability of steady two-dimensional flows through
  imperfect velocity-impulse diagrams'', {\it J. Fluid Mech.} {\bf
    706}, 323--350, (2012).

\bibitem{dr04} P.~G.~Drazin and W.~H.~Reid, {\em Hydrodynamic
    Stability}, (Second edition), Cambridge University Press, 2004.


\bibitem{mb02}
A.~J.~Majda  and A.~L.~Bertozzi, {\em Vorticity and Incompressible Flow},
Cambridge University Press, (2002).

\bibitem{ch09}
S.~Childress, {\em An Introduction to Theoretical Fluid Mechanics},
American Mathematical Society, (2009).

\bibitem{wmz06}
J.-Z.~Wu, H.-Y.~Ma and M.-D.~Zhou, ``Vorticity and Vortex Dynamics'',
Springer, (2006).

\bibitem{t83a}
B.~Turkington, ``On Steady vortex flow in two dimensions. Part I'',
{\em Comm. in Partial Differential Equations} {\bf{8}}, 999--1030,
(1983).

\bibitem{t83b}
B.~Turkington, ``On Steady vortex flow in two dimensions. Part II'',
{\em Comm. in Partial Differential Equations} {\bf{8}}, 1031--1071,
(1983).

\bibitem{gipz09a}
F.~Gallizio, A.~Iollo, B.~Protas, and L.~Zannetti, ``On Continuation
of Inviscid Vortex Patches'', {\em Physica D} {\bf 239}, 190--201, (2010).

\bibitem{sz92}
J.~Sokolowski and J.-P.~Zol\'esio, {\em Introduction to shape
optimization: shape sensitivity analysis}, Springer, (1992).

\bibitem{dz01a}
M.~C.~Delfour and J.-P.~Zol\'esio, ``Shape and Geometries ---
Analysis, Differential Calculus and Optimization'', SIAM, (2001).

\bibitem{hm03}
J.~Haslinger and R.~A.~E.~M\"akinen, {\em Introduction to Shape
Optimization: Theory, Approximation and Computation}, SIAM, (2003).

\bibitem{ss10} S.~Schmidt and V.~Schulz, ``Shape derivatives for
  general objective functions and the incompressible Navier--Stokes
  equations'', {\em Control and Cybernetics} {\bf{39}}, 677--713,
  (2010).

\bibitem{dz04}
M.~C.~Delfour and J.-P.~Zol\'esio, ``Dynamical free boundary problem
for an incompressible potential fluid flow in a time-varying
domain'', {\em Journal of Ill-Posed and Inverse Problems} {\bf{1}},
1--25, (2004).

\bibitem{Pullin81} D.~Pullin, "The nonlinear behavior of a constant
  vorticity layer at a wall", {\em J. Fluid Mech.}{\bf 108}, 401-412,
  (1981).

\bibitem{h95}
W.~Hackbusch, ``Integral Equations: Theory and Numerical Treatment'',
Birkh\"auser, (1995).

\bibitem{p09a} B.~Protas, ``Vortex Design Problem'', {\em Journal of
    Computational and Applied Mathematics} {\bf{236}}, 1926--1946,
  2012.

\bibitem{efm08} A.~Elcrat, B.~Fornberg, and K.~Miller, ``Steady
  axisymmetric vortex flows with swirl and shear'', {\em J. Fluid
    Mechanics} {\bf{613}}, 395--410, (2008).

\bibitem{ph06} D.~S.~Pradeep And F. Hussain, ``Transient growth of
  perturbations in a vortex column'', {\em Journal of Fluid Mechanics}
  {\bf{550}}, 251--288, (2006).

\bibitem{msb12} X.~Mao, S.~J.~Sherwin and H.~M.~Blackburn,
  ``Non-normal dynamics of time-evolving co-rotating vortex pairs''
  {\em Journal of Fluid Mechanics} {\bf{701}}, 430--459, (2012).

\bibitem{em95} A.~Elcrat and K.~G.~Miller, ``Variational Formulas on
  Lipshitz Domains'', {\em Transactions of the AMS} {\bf{347}},
  2669--2678 (1995).
\end{thebibliography}
\end{document}